\documentclass[paper]{ptephy_v1} 

\usepackage{bm}

\usepackage{color}
\usepackage{ulem}

\newcommand{\nc}{\newcommand}           
\nc{\vc}[1]     {\mbox{\boldmath $#1$}} 
\nc{\mapleft}[1]{                       
 \smash{\mathop{                      %
  \hbox to 0.90cm{\rightarrowfill} }\limits_{#1}}}

\nc{\beq}     {\begin{eqnarray}}
\nc{\eeq}    {\end{eqnarray}}
\nc{\bra}       {\langle}               
\nc{\ket}       {\rangle}               
\nc{\bras}[1]   {\langle#1|}            
\nc{\kets}[1]   {|#1\rangle}            
\nc{\del}       {\partial}              

\newcommand{\lw}[1]{\smash{\lower1.75ex\hbox{#1}}}

\nc{\red}[1]    {\textcolor{black}{#1}}  

\nc{\mydraft}	{\setlength{\topmargin}{-1.5cm}}
\mydraft
\begin{document}

\title{
  Cluster configurations in Li isotopes in the variation of multi-bases of the antisymmetrized molecular dynamics
}

\author[1,2]{Takayuki Myo}
\affil[1]{General Education, Faculty of Engineering, Osaka Institute of Technology, Osaka 535-8585, Japan}
\affil[2]{Research Center for Nuclear Physics (RCNP), Osaka University, Ibaraki, Osaka 567-0047, Japan}

\author[3]{Mengjiao Lyu}
\affil[3]{College of Science, Nanjing University of Aeronautics and Astronautics, Nanjing 210016, China}

\author[4]{Qing Zhao}
\affil[4]{School of Science, Huzhou University, Huzhou 313000, Zhejiang, China}

\author[5]{Masahiro Isaka}
\affil[5]{Science Research Center, Hosei University, 2-17-1 Fujimi, Chiyoda, Tokyo 102-8160, Japan}

\author[6]{Niu Wan}
\affil[6]{School of Physics and Optoelectronics, South China University of Technology, Guangzhou 510641, China}

\author[2,7]{Hiroki Takemoto}
\affil[7]{Faculty of Pharmacy, Osaka Medical and Pharmaceutical University, Takatsuki, Osaka 569-1094, Japan} 

\author[2]{Hisashi Horiuchi}

\author[8,9]{Akinobu Dot\'e}
\affil[8]{KEK Theory Center, Institute of Particle and Nuclear Studies (IPNS), High Energy Accelerator Research Organization (KEK), Tsukuba, Ibaraki, 305-0801, Japan}
\affil[9]{J-PARC Branch, KEK Theory Center, IPNS, KEK, Tokai, Ibaraki, 319-1106, Japan}


\begin{abstract}%
  We investigate the cluster configurations in Li isotopes,
  which are described in the optimization of the multi-Slater determinants of the antisymmetrized molecular dynamics.
  Each Slater determinant in the superposition is determined simultaneously in the variation of the total energy.
  The configurations of the excited states are obtained by imposing the orthogonal condition to the ground-state configurations.
  In Li isotopes, various cluster configurations are confirmed and are related to the thresholds of the corresponding cluster emissions.
  For $^5$Li, we predict the $^3$He+$d$ clustering in the excited state as well as the mirror state of $^5$He with $^3$H+$d$. 
  For $^{6-9}$Li, various combinations of the clusters are obtained in the ground and excited states,
  and the superposition of these basis states reproduces the observed energy spectra.
  For $^9$Li, we predict the linear-chain states consisting of various cluster configurations at 10--13 MeV of the excitation energy.
\end{abstract}

\subjectindex{D10, D11, D13}

\maketitle

\section{Introduction}

Clustering phenomena is an important aspect of nuclei \cite{ikeda68,horiuchi12,freer18}.
A typical case is the $\alpha$ cluster state, such as $^8$Be and the Hoyle state of $^{12}$C ($0^+_2$).
These cluster states are often observed near the threshold energy of the $\alpha$ particle emission
and this property is known as the ``threshold rule'' \cite{ikeda68}.
The Ikeda diagram indicates the variety of nuclear clustering in the $N\alpha$ system systematically according to the threshold rule.
Nuclear clustering is also expected in unstable nuclei having excess protons and neutrons with their weak bindings and excitations \cite{horiuchi12}.

Theoretically, it is desirable to investigate the existence of the nuclear cluster states without any assumption.
The antisymmetrized molecular dynamics (AMD) is the possible nuclear model for this purpose \cite{kanada03,kimura16}.
In AMD, the nucleon wave function has a Gaussian wave packet with a centroid position.
The centroid parameters control the formations of the cluster and mean-field states in nuclei
and they can be determined by the variational principle of the total nuclear energy.

In AMD, the extension to the multi-Slater determinants, namely the multi-bases, is straightforward by applying the generator coordinate method (GCM).
However, the problem of multi-bases is the construction of the optimal AMD basis states to be superposed,
which is not only for the ground state but also for the excited states of nuclei.
In the previous work \cite{myo23b}, we proposed a promising method to determine the optimal multi-Slater determinants in AMD.
\red{
  A similar attempt has been discussed so far in the mean-field framework of nuclei \cite{faessler71,ogawa11,pillet17} including the recent one \cite{matsumoto23} and also in the shell-model approach \cite{shimizu22}.
  }

\red{
We consider the variation of the total energy of the superposed AMD wave function
by extending the cooling method (the imaginary-time evolution) for a single AMD basis state to the multi-basis case.
This is the cooling for the multi-AMD bases, namely the multiple cooling, and then we simply abbreviate this method as the ``multicool method''
\cite{myo23b,tian24,cheng24}.
In this method, we can determine all the parameters in the multi-basis states simultaneously to minimize the total energy of the superposed wave function.
}

We use this multicool method to obtain the configurations of the nuclear ground state.
We further developed the method to generate the configurations of the excited states by imposing the orthogonal condition to the ground-state configurations.
Recently, this multicool method has been combined with the control neural network and
is applied to the $\Lambda$ hypernuclei of Be isotopes \cite{tian24} and the $3\alpha$ resonances of $^{12}$C \cite{cheng24}.

In the previous work of $^{10}$Be \cite{myo23b}, we describe the ground and excited states in the multicool method.
Among the cluster models using the same Hamiltonian, the multicool results provide the lowest energies of the states.
We obtain not only the shell-like configuration but also various cluster configurations of $^9$Be+$n$, $^8$Be+$2n$, $^7$Li+$t$, $^6$He+$\alpha$, and $^5$He+$^5$He, in which the unbound subsystem such as $^5$He can be a constituent of a clustering system.
These cluster configurations are obtained automatically and simultaneously in the variation of total energy with the multi-AMD basis states.
They also contribute to forming the linear-chain structure in the excited states of $^{10}$Be.

The appearance of the cluster configurations can be related to the thresholds of the corresponding cluster emissions.
Based on the results of $^{10}$Be, it is interesting to investigate the coexistence of the various cluster configurations in nuclei in the multicool method.

In this study, we focus on the Li isotopes.
In Li isotopes, many kinds of thresholds of the cluster emissions exist in the low-excitation energy region involving the $\alpha$ and $t$ clusters.
In the case of $^9$Li, the isotone of $^{10}$Be, the thresholds of $^8$Li+$n$, $^7$Li+$2n$, $^6$He+$t$, and $\alpha$+$t$+$n$+$n$ are located experimentally 
below 10 MeV of the excitation energy \cite{tilley04,tunl,nndc}.
These cluster configurations are expected to contribute to the structure of $^9$Li as well as the shell-like one, similar to $^{10}$Be.
In addition, for $^5$Li, one usually considers the mean-field structure consisting of $\alpha$+$p$ in the ground state region.
In the high excitation energy region, there is a threshold of $^3$He+$d$ and the corresponding cluster states are expected to exist in the excited states.
This indicates the variety of the clustering in nuclei.

Hence it is worthwhile to investigate the appearance of the possible cluster configurations in Li isotopes in the multicool method with AMD.
This study is related to the examination of the threshold rule for unstable nuclei.
In connection to the $\alpha$ cluster ($^4$He) emergence in Li isotopes,
we apply the multicool method to $^4$He and discuss the excited $0^+$ states, the energies of which are close to the $t$+$p$ threshold.

We demonstrate that the present multicool method is useful to generate the optimal AMD configurations to be superposed in the GCM calculation.
Finally, we confirm the reliability of the multicool method to investigate the nuclear structure
from the aspect of the multi-configurations with the clustering effect.

In Sec.~\ref{sec:method}, we explain the multicool method for variation of the multi-Slater determinants in the AMD wave functions.
In Sec.~\ref{sec:result}, we discuss the structures of Li isotopes.
In Sec.~\ref{sec:summary}, we summarize the present work.

\section{Method}\label{sec:method}

\subsection{Hamiltonian}\label{sec:ham}
We use the Hamiltonian with a two-body nucleon-nucleon interaction for mass number $A$:
\begin{eqnarray}
    H
&=& \sum_i^{A} t_i - T_{\rm c.m.} + \sum_{i<j}^{A} v_{ij},
    \label{eq:ham}
    \\
    v_{ij}
&=& v_{ij}^{\rm central} + v_{ij}^{\rm LS} + v_{ij}^{\rm Coulomb}.
\end{eqnarray}
Here $t_i$ and $T_{\rm c.m.}$ are the kinetic energies of each nucleon and the center-of-mass (c.m.), respectively.
Following the previous works of $^{10}$Be \cite{myo23b,itagaki00,ito06,suhara10},
we use the effective nucleon-nucleon interaction $v_{ij}$ of the Volkov No.2 central force with the parameters  $(W,M,B,H)=(0.4, 0.6, 0.125, 0.125)$, 
the G3RS LS force with 1600 MeV of the strength, and the point Coulomb force.
This effective interaction is often used in the description of light nuclei with cluster models by adjusting $M(=1-W)$ and the strength of LS force.

\subsection{Antisymmetrized molecular dynamics}\label{sec:AMD}

The AMD wave function $\Phi_{\rm AMD}$ for the $A$-nucleon system is expressed in a single Slater determinant:
\begin{eqnarray}
\Phi_{\rm AMD}
&=& \frac{1}{\sqrt{A!}} {\rm det} \biggl\{ \prod_{i=1}^A \phi_i(\bm{r}_i) \biggr\}~,
\label{eq:AMD}
\\
\phi_i(\bm{r})&=&\left(\frac{2\nu}{\pi}\right)^{3/4} e^{-\nu(\bm{r}-\bm{Z}_i)^2} \chi_{\sigma,i}\, \chi_{\tau,i},
\\
\chi_{\sigma,i} &=& \alpha_i \kets{\uparrow} + \beta_i \kets{\downarrow}.
\label{eq:Gauss}
\end{eqnarray}
The single-nucleon wave function $\phi_i(\bm{r})$ has a Gaussian wave packet with a range parameter $\nu$
and the centroid position $\bm{Z}_i$ with the index $i=1,\cdots,A$.
We set $\sum_{i=1}^A \bm{Z}_i={\bf 0}$.
We use $\nu=0.235$ fm$^{-2}$, which is the same value as used in the previous works \cite{myo23b,itagaki00,suhara10}.
For deuteron, the single AMD configuration gives the energy of $-0.67$ MeV,
while the exact solution reproduces the experimental energy \cite{suhara10}.
The energies of the $(0s)^A$ configurations for $^3$H, $^3$He, and $^4$He are $-6.88$ MeV, $-6.09$ MeV, and $-27.59$ MeV, respectively.

The spin wave function $\chi_{\sigma}$ is a mixed state of the up ($\uparrow$) and down ($\downarrow$) components for the $z$ direction
and the isospin one $\chi_{\tau}$ is a proton or neutron.
We define $X_i :=\{\bm{Z}_i,\alpha_i,\beta_i\}$ for the set of variational parameters. 

The energy minimization is carried out by solving the cooling equation
for total energy $E^\pm_{\rm AMD}$ with parity projection $P^\pm$ using the arbitrary positive number $\mu$,
and the parameters $\{X_i\}$ are determined by the following equations:
\begin{equation}
  \begin{split}
    \Phi^\pm_{\rm AMD}&= P^\pm \Phi_{\rm AMD},
    \\
    E^\pm_{\rm AMD} &= \dfrac{ \bra \Phi^\pm_{\rm AMD}|H| \Phi^\pm_{\rm AMD} \ket }{\bra \Phi^\pm_{\rm AMD}| \Phi^\pm_{\rm AMD} \ket},
    \\
    \dfrac{{\rm d} X_i}{{\rm d} t}&= - \mu \dfrac{\partial E^\pm_{\rm AMD}}{ \partial X_i^*},\quad \mbox{and c.c}.
  \end{split}
  \label{eq:cooling}
\end{equation}

The angular-momentum projection with the operator $P^J_{MK}$
is carried out for the angular momentum $J$ with quantum numbers of $M$ and $K$, and the parity ($\pm$):
\begin{eqnarray}
  \begin{split}
  \Psi^{J^\pm}_{MK,{\rm AMD}}
  &= P^J_{MK}P^{\pm} \Phi_{\rm AMD}.
  \end{split}
  \label{eq:projection}
\end{eqnarray}

The AMD wave function is extendable to the multi-configuration with GCM.
The GCM wave function $\Psi_{\rm GCM}$ is a superposition of the projected AMD basis states denoted as $\Psi_n$ simply 
with the basis index of $n=1,\cdots,N$:
\begin{eqnarray}
  \begin{split}
   \Psi_{\rm GCM}
&= \sum_{n=1}^N C_n\,  \Psi_n \,,
   \\
   H_{mn}
&= \langle\Psi_{m} | H |\Psi_{n}\rangle \,,
   \quad
   N_{mn}
=  \langle\Psi_{m} | \Psi_{n} \rangle \,,
   \\
   E_{\rm GCM} &= \dfrac{ \bra \Psi_{\rm GCM} |H| \Psi_{\rm GCM} \ket }{\bra \Psi_{\rm GCM}| \Psi_{\rm GCM} \ket}, 
  \end{split}
   \label{eq:linear}
\end{eqnarray}
where the labels $m$ and $n$ represent the set of the quantum numbers of the projected AMD basis states in Eq.~(\ref{eq:projection}).
The Hamiltonian and norm matrix elements are given as $H_{mn}$ and $N_{mn}$, respectively.
From the variational principle, the generalized eigenvalue problem (Hill-Wheeler equation) is solved to obtain $\Psi_{\rm GCM}$ and $E_{\rm GCM}$:
\begin{eqnarray}
   \sum_{n=1}^N \bigl( H_{mn} - E_{\rm GCM} N_{mn} \bigr) C_n &=& 0.
   \label{eq:eigen}
\end{eqnarray}
\subsection{Variation of multi-AMD basis states}\label{sec:multi}

In the previous study \cite{myo23b}, we extended the cooling equation in Eq.~(\ref{eq:cooling}) to optimize the multi-AMD basis states
in the energy variation of the total system.
We propose the extended cooling equation, which is called the ''multicool method''.

We express the total wave function $\Phi$ in the linear combination of the intrinsic AMD basis states $\{\Phi_n\}$ with a number $N$
and the total energy $E$ is given as
\begin{equation}
  \begin{split}
   \Phi&= \sum_{n=1}^N C_n\,  \Phi_n , \qquad
   E    = \dfrac{ \bra \Phi |H| \Phi \ket }{\bra \Phi | \Phi \ket}.
  \end{split}
  \label{eq:multi}
\end{equation}
In this study, the parity projection is always performed and we omit the notation of parity $(\pm)$.

The AMD configurations have variational parameters of $X_{n,i}=\{\bm{Z}_{n,i},\alpha_{n,i},\beta_{n,i},C_n\}$.
The cooling equation to minimize the total energy $E$ is expressed as 
\begin{equation}
  \begin{split}
   \dfrac{{\rm d} X_{n,i}}{{\rm d} t}&= - \mu \dfrac{\partial E}{ \partial X_{n,i}^*},\quad \mbox{and c.c}.
  \end{split}
  \label{eq:multi_eq}
\end{equation}
By solving this equation numerically, the total energy $E$ and $\{X_{n,i}\}$ are obtained for the nuclear ground state.
After the angular-momentum projection of the basis states $\{\Phi_n\}$, one solves the equation in Eq.~(\ref{eq:eigen}).

We further introduce the method to generate the configurations of the excited states, which are favored to be orthogonal to the ground-state configurations.
We employ the orthogonal projection method conventionally used in the orthogonality condition model \cite{aoyama01,myo14}.

\red{
  We first obtain the ground-state configurations $\{\Phi_n\}$ in Eq.~(\ref{eq:multi}) in the intrinsic frame.
  When the configuration $\Phi_n$ is deformed to a specific direction,
  the configuration orthogonal to $\Phi_n$ can be deformed to another direction, but
  these configurations might have an overlap after the angular momentum projection. 
  To avoid this situation, we add the rotated configurations of $\{\Phi_n\}$ in the ground state.
  In this study, we adopt two kinds of rotations; $(x,y,z)\to(z,x,y)$ and $(x,y,z)\to(y,z,x)$ according to the previous work \cite{myo23b}.
  Finally, we define the ground-state configurations $\{\Phi_{c}\}$ with the index $c=1,\cdots,3N$,
  and consider the orthogonality to $\{\Phi_{c}\}$ to generate the excited states.
}

Next, we define the pseudo potential $V_\lambda$ using $\{\Phi_{c}\}$ in the projection form with a strength $\lambda$ and add $V_\lambda$ to the Hamiltonian: 
\begin{equation}
  \begin{split}
    H_\lambda &= H + V_\lambda,\quad
    V_\lambda = \lambda \sum_{c=1}^{3N} \kets{\Phi_{c}}\bras{\Phi_{c}},
    \\
    \Phi_{\lambda} &= \sum_{n=1}^N C_{\lambda,n} \Phi_{\lambda,n},\quad
    E_{\lambda} = \dfrac{ \bra \Phi_{\lambda} |H_{\lambda}| \Phi_{\lambda} \ket }{\bra \Phi_{\lambda} | \Phi_{\lambda} \ket}.
  \end{split}
  \label{eq:PSE}
\end{equation}
Using a specific value of $\lambda$, we perform the variation of the total energy $E_{\lambda}$ in Eq. (\ref{eq:multi_eq})
and determine the basis states $\{\Phi_{\lambda,n}\}$.
We start the calculation with a small value of $\lambda$ and gradually increase $\lambda$.
When $\lambda$ becomes large, $\Phi_{\lambda}$ can be orthogonal to the configurations $\{\Phi_c\}$. 
It is noted that $\Phi_{\lambda}$ is orthogonal to $\{\Phi_c\}$, but each of the configuration $\Phi_{\lambda,n}$ is not
imposed to be orthogonal to $\{\Phi_c\}$.
After the convergence of the solutions in Eq. (\ref{eq:multi_eq}),
we obtain the total energy $E_{\lambda}$ omitting the contribution of the pseudo potential.
We use this technique to produce the configurations of the excited states $\{\Phi_{\lambda,n}\}$ in the multicool method.

We explain the whole procedure of the multicool method for a nucleus with a mass number $A$.
\begin{enumerate}
\itemsep=3mm
\item[(a)]
  We prepare the initial AMD basis states randomly
  with the parameters $\{X_{n,i}\}$ for the basis index $n=1,\cdots,N$ and the particle index $i=1,\cdots,A$.
  We set the initial coefficients $\{C_n\}$ solving the eigenvalue problem of the intrinsic Hamiltonian matrix.

\item[(b)]
  Solving the multicool equation in Eq. (\ref{eq:multi_eq}), we obtain the basis states $\{\Phi_{n}\}$ for the ground state.
  We define $\{\Phi_{c}\}$ with $c=1,\cdots,3N$ using $\{\Phi_n\}$ including their rotations.
  
\item[(c)]
  Using the pseudo potential with the strength $\lambda$ in Eq.~(\ref{eq:PSE}),
  we solve the multicool equation and obtain the configurations $\{\Phi_{\lambda,n}\}$ of the excited states.

\item[(d)]
  We superpose $\{\Phi_n\}$ and $\{\Phi_{\lambda,n}\}$ with various values of $\lambda$ and
  solve the eigenvalue problem of the Hamiltonian matrix with the projection in Eq. (\ref{eq:eigen}). 
\end{enumerate}

In the multicool calculation, the number of the basis states is typically around $N=15-20$ in Eq.~(\ref{eq:multi}). 
In the variation of the total energy, we employ two conditions on the intrinsic spin of nucleons according to the previous study \cite{myo23b};
one is that the spin directions are fixed with spin-up ($\alpha_i=1,\beta_i=0$) and spin-down ($\alpha_i=0,\beta_i=1$) in Eq.~(\ref{eq:Gauss}) (spin-fix).
The other is that the spin directions are optimized during the energy variation (spin-free).
In the GCM wave function, we superpose the basis states obtained in two conditions. 
In the present study, the total basis number is typically at most about 500 to 600 in Li isotopes.

\red{
  It is noted that if we do not use the pseudo potential $V_\lambda$, the energy variation is carried out for the ground state
  and we obtain only the ground-state configurations $\{\Phi_n\}$ in Eq.~(\ref{eq:multi}).
  Even if we increase the basis number $N$, it is still difficult to obtain the configurations, which are suitable to describe the excited states.
}

\red{
  We remark the possible advantage of the multicool method; 
  one can generate the optimal basis states simultaneously for nuclei according to the energy variational principle
  and this process might be faster than a Monte-Carlo search over the possible Slater determinants \cite{shimizu22}.
  On the other hand, the application of the present method is limited to the light-mass region of nuclei,
  because the calculation cost depends on the number of degrees of freedom in each Slater determinant,
  which increases with a mass number, as well as the number of basis states.
}

\section{Results}\label{sec:result}

\subsection{multicool calculation for $^9$Li}\label{sec:calculation}

We demonstrate one case of the multicool calculation; We show the results of $^9$Li for the intrinsic negative parity state with the basis number $N=18$ and the spin-fix case.
A similar analysis is done in the previous study of $^{10}$Be \cite{myo23b}.

\begin{figure}[b]
\centering
\includegraphics[width=6.8cm,clip]{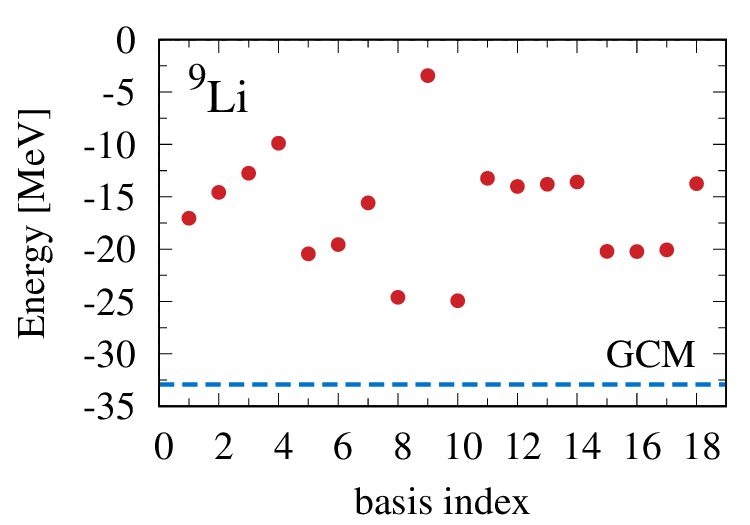}\hspace*{0.5cm}
\includegraphics[width=6.8cm,clip]{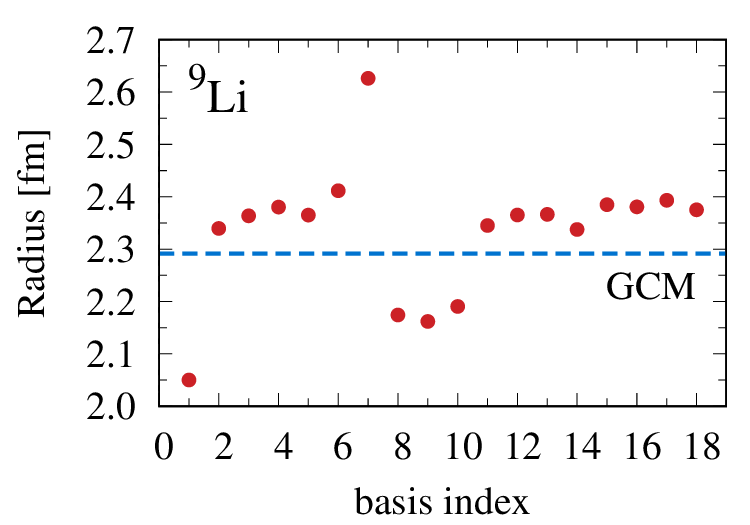}\\
\includegraphics[width=6.8cm,clip]{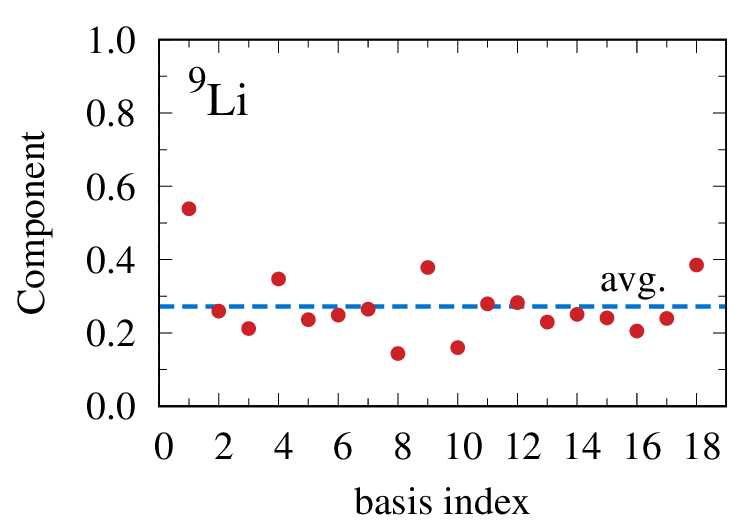}
\caption{
  Configurations of $^9$Li in the spin-fix case for the intrinsic negative parity state.
  The horizontal axis is the basis index $n$ in the multicool method.
  The red dots mean the total energies in units of MeV (upper left), matter radii in units of fm (upper right),
  and the components $| \langle \Phi_n| \Phi \rangle |^2$ (lower) of each configuration.
  The blue dashed lines represent the energy and radius of the intrinsic GCM wave function and the average of the components.}
\label{fig:ene_9_demo}
\end{figure}

In Fig.~\ref{fig:ene_9_demo}, we show the total energies (upper left panel), matter radii (upper right panel), and the components (lower panel) of the configurations of $^9$Li, 
in which the component with the basis index $n$ is defined as $|\langle \Phi_n|\Phi \rangle|^2$ in Eq.~(\ref{eq:multi}).
It is noted that the AMD basis states $\{\Phi_n\}$ are non-orthogonal to each other.
The total energy and the radius of the intrinsic GCM wave function $\Phi$ are shown with the blue dashed lines.
The obtained configurations are optimized to describe the ground state of $^9$Li.

In the results, the energies of the individual configurations are distributed widely from $-3$ MeV to $-25$ MeV and the GCM energy is $-33$ MeV with the energy gain of 8 MeV.
For radius, the values are also distributed widely from 2.05 fm to 2.62 fm and the GCM result is 2.29 fm.
For the components, all configurations contribute to the total wave function indicating the importance of the configuration mixing
and their average is 0.27.

\begin{figure}[t]
\centering
\includegraphics[width=13.0cm,clip]{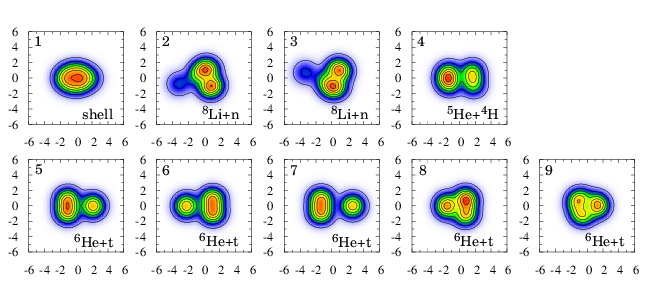} 
\caption{
  Intrinsic density distributions of the configurations of $^9$Li in the spin-fix case.
  Units of densities and axes are fm$^{-3}$ and fm, respectively.
  The left-top number in each panel means the basis index in Fig. \ref{fig:ene_9_demo}.
}
\label{fig:density_9_demo}
\end{figure}

In Fig.~\ref{fig:density_9_demo}, we show the intrinsic density distributions of the nine representative configurations of $^9$Li,
which are selected among the 18 basis states, and the basis index is assigned from 1 to 9 in Fig.~\ref{fig:ene_9_demo}.
Here, the integration of the distribution in three-dimensional space gives a mass number.
We adjust the direction of the longest distribution to be the horizontal axis.

From the density distributions, we confirm the variety of the $^9$Li configurations.
The 1st configuration shows the compact shell-like structure with the smallest radius of 2.05 fm among the basis states.
The 2nd and 3rd configurations show the $^8$Li+$n$ cluster structure where $^8$Li shows the spatially compact cluster structure consisting of $^5$He+$t$.
The 4th configuration shows the $^5$He+$^4$H cluster structure, even though both clusters are unbound in the isolated system.
Their relative distance is short, indicating that the interaction between clusters is not weak.
The 5th, 6th, and 7th configurations show the $^6$He+$t$ cluster structure and
the relative distance between clusters depends on the configurations and the 7th configuration shows the largest radius of 2.62 fm among the basis states.
The 8th and 9th configurations commonly show the compact $^6$He+$t$ structure with a small radius of around 2.2 fm as shown in Fig. \ref{fig:ene_9_demo}.

Various cluster configurations are confirmed in $^9$Li in the multicool method.
It is noted that the appearance of these configurations is related to the existence of the thresholds corresponding to the cluster emissions.
In the experiments of $^9$Li, the threshold energies of the $^8$Li+$n$, $^7$Li+$2n$, $^6$He+$t$, and $\alpha$+$t$+$2n$ separations
are 4.1 MeV, 6.1 MeV, 7.6 MeV, and 8.6 MeV, respectively, measured from the ground state \cite{tilley04,tunl,nndc}.
Considering the above cluster thresholds, there is an analysis of the ground-state properties of $^9$Li in the $\alpha$+$t$+$2n$ cluster model \cite{varga95}.

\begin{figure}[t]
\centering
\includegraphics[width=6.8cm,clip]{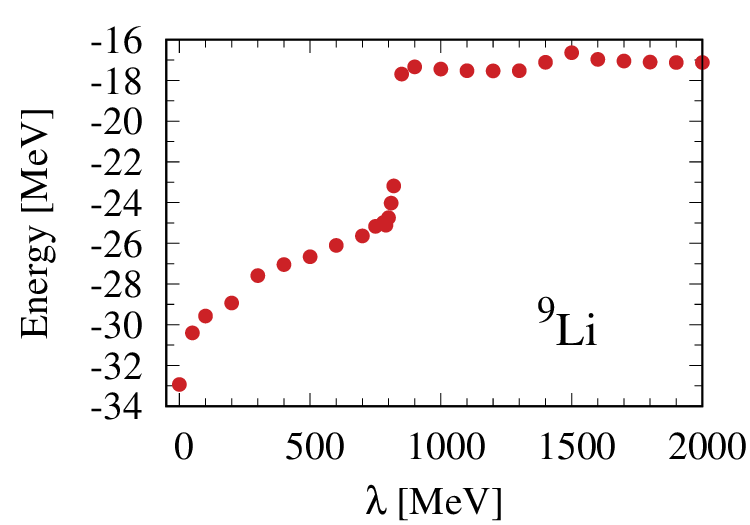}~~~
\includegraphics[width=6.8cm,clip]{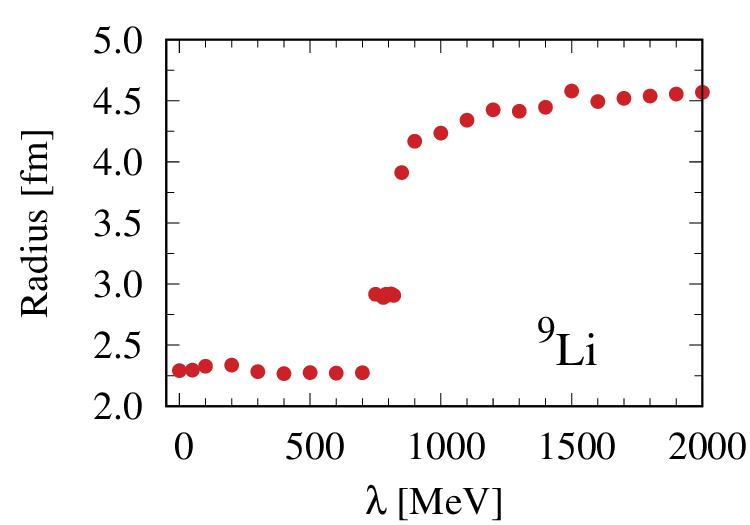}
\caption{
  Intrinsic energy (left) and radius (right) of the total wave function of $^9$Li with negative parity in the spin-fix case.
  The strength $\lambda$ of the pseudo potential changes.}
\label{fig:ene_lam}
\end{figure}

\begin{figure}[t]
\centering
\includegraphics[width=15.3cm,clip]{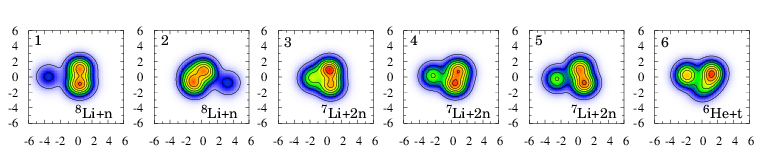} 
\caption{
  Intrinsic density distributions of the configurations of $^9$Li in the spin-fix case
  using the pseudo potential with the strength of $\lambda=500$ MeV.
  The number in each panel means the basis index in order.
}
\label{fig:density_9_pse500}
\end{figure}

Next, we introduce the pseudo potential to generate the configurations of the excited states in the multicool method.
In Fig. \ref{fig:ene_lam}, the total energy (left panel) and the matter radius (right panel) of the intrinsic negative parity state of $^9$Li
with GCM are shown as the strength of $\lambda$ increases in the pseudo potential in Eq. (\ref{eq:PSE}).
Here, at $\lambda=0$, we plot the ground-state values as shown in Fig. \ref{fig:ene_9_demo}.
It is found that as $\lambda$ increases, the system is gradually excited and the radius also changes.
At around $\lambda=800$ MeV, the energy and radius suddenly increase and this behavior suggests the change of the internal structure of $^9$Li,
which is discussed in terms of the density distributions of the AMD configurations.
At around $\lambda=2000$ MeV, the energy becomes stable to $-17$ MeV and the radius reaches 4.5 fm.

\begin{figure}[t]
\centering
\includegraphics[width=13.0cm,clip]{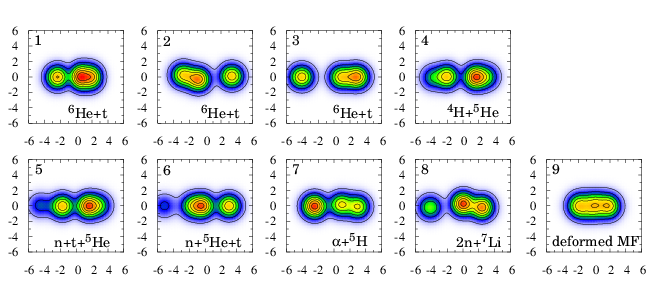} 
\caption{
  The same as Fig. \ref{fig:density_9_pse500}, but with the strength of $\lambda=800$ MeV.
}
\label{fig:density_9_pse800}
\end{figure}
\begin{figure}[t]
\centering
\includegraphics[width=15.3cm,clip]{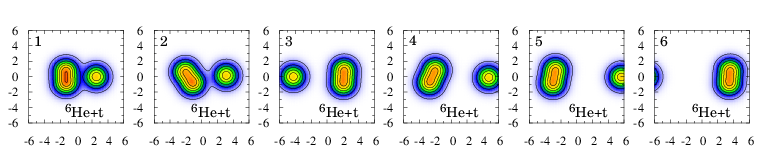} 
\caption{
  The same as Fig. \ref{fig:density_9_pse500}, but with the strength of $\lambda=1600$ MeV.
}
\label{fig:density_9_pse1600}
\end{figure}

We show the intrinsic density distributions with several values of $\lambda$.
In Fig. \ref{fig:density_9_pse500}, we show the results with $\lambda=500$ MeV, in which the six representative configurations are selected putting the basis index in each panel in order.
The 1st and 2nd configurations show the $^8$Li+$n$ configuration where $^8$Li still shows the compact cluster structure and the positions of the last neutron are different with respect to $^8$Li. 
The 3rd, 4th, and 5th configurations show the $^7$Li+$2n$ configuration, where $^7$Li shows the compact cluster structure consisting of $^4$He+$t$
and the clustering of $2n$ depends on the configurations.
The 6th configuration shows the $^6$He+$t$ cluster where $^6$He is inclined with respect to the vertical direction,
which is different from the ground-state case shown in Fig. \ref{fig:ene_9_demo}.
From the results, the $^7$Li+$2n$ configuration is newly obtained by exciting $^9$Li with the pseudo potential.

In Fig. \ref{fig:density_9_pse800}, we show the density distributions with $\lambda=800$ MeV, in which the nine representative configurations are selected with the basis index.
From the property of these configurations, they commonly show the elongated linear-chain structure to the horizontal axis,
which is similar behavior to the results of the excited state of $^{10}$Be \cite{myo23b}.
In the configurations, there is a variety of cluster structures; $^6$He+$t$ (1st, 2nd, and 3rd),
$^4$H+$^5$He (4th), $n$+$t$+$^5$He (5th), $n$+$^5$He+$t$ (6th), $\alpha$+$^5$H (7th), $2n$+$^7$Li (8th),
and as well as the deformed mean-field (MF) type (9th).
For $^6$He+$t$, the relative distance between clusters and the intrinsic shape of $^6$He depend on the configurations.
It is an interesting aspect that the unbound nuclei of $^5$He and $^{4,5}$H are involved to be the constituents of a clustering system.

In Fig. \ref{fig:density_9_pse1600}, we show the results with $\lambda=1600$ MeV, in which the six representative configurations are shown.
The configurations mainly show the $^6$He+$t$ structure with various relative distances between clusters
and the elongated direction of $^6$He in $^9$Li is different from the linear-chain structure obtained at $\lambda=800$ MeV as shown in Fig. \ref{fig:density_9_pse800}.

In the final calculation, we superpose all the basis states obtained with the spin-fix and spin-free conditions in the multicool method and perform the angular-momentum projection. 
We solve the eigenvalue problem to obtain the energy levels in the GCM calculation, which are discussed later with the results of other Li isotopes.


\subsection{Energy and radius}\label{sec:result}

We generate the AMD basis states in the multicool method without and with the pseudo potential for spin-fix and spin-free cases and perform the GCM calculation
for Li isotopes.
In Tables \ref{tab:ene_radius1} and \ref{tab:ene_radius2}, we show the results of the total energies and radii
of Li isotopes for mainly the ground state and also $s$-shell nuclei of $^3$H, $^3$He, $^4$He and $^6$He for references \cite{myo23b}.
For unbound states, we adopt the bound-state approximation.

It is found that the total energies and the charge radii of the ground states of nuclei reproduce the experimental values systematically
with the present Hamiltonian in the multicool method.
For $^8$Li and $^9$Li, one can see the underbinding by about $3-4$ MeV.
This can be related to the strength of the LS force because the contributions of LS force increase from $^7$Li ($-2.1$ MeV) to $^9$Li ($-10.5$ MeV).
When we strengthen the LS force with 2000 MeV \cite{itagaki00,furumoto13}, total energies of $^7$Li, $^8$Li, and $^9$Li are $-39.6$ MeV, $-39.7$ MeV, and -$44.6$ MeV, respectively,
which become close to the experimental values, although the energy of $^{10}$Be becomes overbound by 1.8 MeV with respect to the experimental value \cite{myo23b}.
In the next sections, we discuss in detail the results of each of Li isotopes including the excited states.

\begin{table}[t]
\begin{center}
  \caption{
    Total energies and radii of matter, proton, neutron, and charge ($r_{\rm m}$, $r_{\rm p}$, $r_{\rm n}$, and $r_{\rm ch}$) 
    of $^3$H, $^3$He, $^4$He, and $^6$He in the multicool calculation.
    The values with square brackets are the experimental energies and charge radii \cite{tunl,sick08,mueller07}.
    Units of energy and radius are MeV and fm, respectively.}
\label{tab:ene_radius1} 
\renewcommand{\arraystretch}{1.1}
\begin{tabular}{llllllllllll}
\noalign{\hrule height 0.5pt}
                &~~Energy~~            &&~$r_{\rm m}$  &&~$r_{\rm p}$ &&~$r_{\rm n}$ &&~$r_{\rm ch}$ \\
\noalign{\hrule height 0.5pt}
$^3$H  ($1/2^+$) & ~$-8.37$ [$-8.48$]  &&  1.67 &&  1.61  &&  1.71  && 1.77 [1.755(86)] \\
$^3$He ($1/2^+$) & ~$-7.68$ [$-7.72$]  &&  1.69 &&  1.73  &&  1.62  && 1.91 [1.959(30)] \\
$^4$He ($0^+_1$) & $-28.90$ [$-28.30$] &&  1.53 &&  1.53  &&  1.52  && 1.75 [1.681(4)] \\
$^4$He ($0^+_2$) & ~$-7.00$ [$-8.09$]  &&  3.75 &&  4.40  &&  2.95  && 4.48             \\
$^6$He ($0^+$)   & $-29.20$ [$-29.27$] &&  2.38 &&  1.88  &&  2.59  && 2.04 [2.068(11)] \\
\noalign{\hrule height 0.5pt}
\end{tabular}
\end{center}
\end{table}

\begin{table}[t]
\begin{center}
  \caption{
    Total energies and radii of matter, proton, neutron, and charge ($r_{\rm m}$, $r_{\rm p}$, $r_{\rm n}$, and $r_{\rm ch}$) 
    of the ground states of Li isotopes in the multicool calculation.
    The values with square brackets are the experimental energies and charge radii \cite{nortershauser11}.
    Units of energy and radius are MeV and fm, respectively.}
\label{tab:ene_radius2} 
\renewcommand{\arraystretch}{1.1}
\begin{tabular}{llllllllllll}
\noalign{\hrule height 0.5pt}
                &~~Energy~~           &&~$r_{\rm m}$  &&~$r_{\rm p}$ &&~$r_{\rm n}$ &&~$r_{\rm ch}$ \\
\noalign{\hrule height 0.5pt}
$^5$Li ($3/2^-$) &  $-26.87$ [$-26.61$] &&  ~~-- && ~~-- && ~~-- && ~~-- \\
$^5$Li ($1/2^-$) &  $-25.34$ [$-25.12$] &&  ~~-- && ~~-- && ~~-- && ~~-- \\
$^6$Li ($1^+$)   &  $-31.41$ [$-32.00$] &&  2.31 && 2.31 && 2.30 && 2.46 [2.59(4)] \\
$^7$Li ($3/2^-$) &  $-38.99$ [$-39.25$] &&  2.39 && 2.30 && 2.46 && 2.44 [2.44(4)] \\
$^8$Li ($2^+$)   &  $-38.07$ [$-41.28$] &&  2.33 && 2.16 && 2.42 && 2.30 [2.34(5)] \\
$^9$Li ($3/2^-$) &  $-41.55$ [$-45.34$] &&  2.29 && 2.06 && 2.40 && 2.20 [2.25(5)] \\
\noalign{\hrule height 0.5pt}
\end{tabular}
\end{center}
\end{table}

\subsection{$^4$He}\label{sec:result4}

\begin{figure}[t]
\begin{flushleft}
\centering
\includegraphics[width=5.6cm,clip]{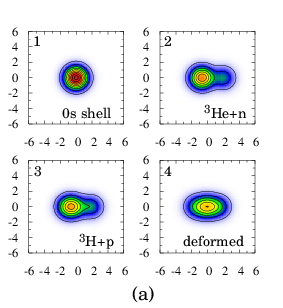}~~~~~~
\includegraphics[width=8.2cm,clip]{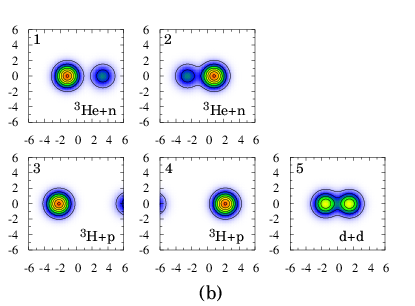}
\end{flushleft}
\caption{
  Intrinsic density distributions of the representative configurations of $^{4}$He for positive parity in the spin-fix case.
  (a) four panels are the configurations in the ground state.
  (b) five panels are the configurations in the excited state.
  Units of densities and axes are fm$^{-3}$ and fm, respectively.
}
\label{fig:density_4}
\end{figure}

\begin{figure}[t]
\centering
\includegraphics[width=7.5cm,clip]{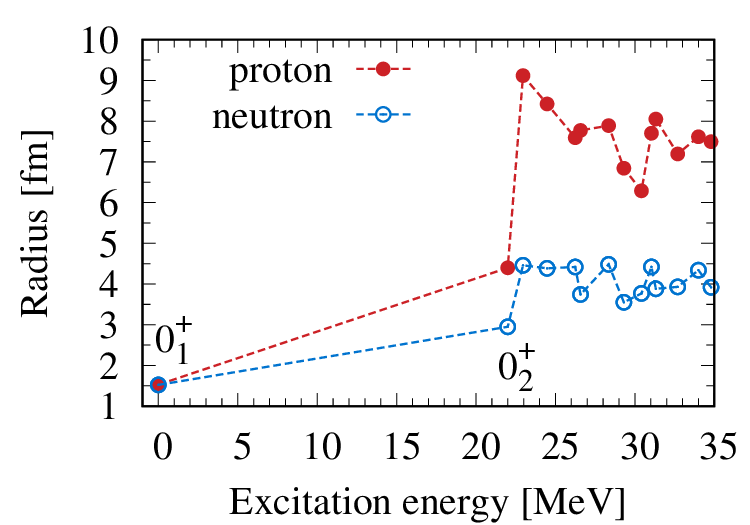}
\caption{
  Radii of protons (red solid circles) and neutrons (blue open circles) of $^4$He ($0^+$) as functions of the excitation energy. Units are fm.
}
\label{fig:radius_4}
\end{figure}

We start to discuss the results of $^4$He in the multicool method
because the $\alpha$ cluster often appears in the configurations of Li isotopes in the present calculations.
In Fig. \ref{fig:density_4} (a), we show the intrinsic density distribution of $^4$He for the ground state with positive parity in the four panels.
In the ground $0^+$ state, the spherical $0s$-shell configuration (1st panel) is dominant with a weight of 98\%.
Additionally, $^3$He+$n$ and $^3$H+$p$ configurations (2nd and 3rd) are also confirmed with a small relative distance and their weight is about 80\%.
Furthermore, the deformed configuration for both protons and neutrons (4th) is obtained with a weight of about 85\%.
Hence the breaking of the $0s$-shell configuration is described in the multicool method and their effect on the total energy is estimated as 1.3 MeV
as shown in Table \ref{tab:ene_radius1} in comparison with $-27.6$ MeV with the $0s$-shell case.

In the five panels of Fig. \ref{fig:density_4} (b), we show the representative configurations of the excited state obtained using the pseudo potential,
where the strength $\lambda$ is taken to generate the configurations largely different from the ground-state ones in the density distribution.
In the results, the $^3$H+$p$ and $^3$He+$n$ configurations are obtained with various relative distances as well as the $d$+$d$ configuration.
The $^3$H+$p$ configuration mainly contributes to the $0^+_2$ state of $^4$He in the GCM calculation.
It looks that the density of $^3$H shows the $0s$-shell configuration giving the total energy of $-6.9$ MeV in the present Hamiltonian.
Experimentally, the $^3$H+$p$ is the lowest threshold of the particle emission.

In the present calculation, the $0^+_2$ state is obtained with a total energy of $-7.0$ MeV, and the excitation energy $E_x=21.9$ MeV, which is close to the threshold energy of $^3$H+$p$ with the $0s$-shell configuration of $^3$H.
Experimentally, the $0^+_2$ state is observed at $E_x=21.1$ MeV, above 0.40 MeV of the $^3$H+$p$ threshold energy, and the present results are consistent with the experimental situation. 
From Table \ref{tab:ene_radius1}, the proton radius is larger than the neutron radius in the $0^+_2$ state.
The maximum probability of the single AMD configuration is 41 \% with a $^3$H+$p$ configuration
and the same cluster configuration with various cluster distances are mixed in this state.
\red{
  The pioneering works on the property of the $0^+_2$ state are performed in the ab-initio four-body calculations \cite{hiyama04,horiuchi08}
  and also in the recent Gamow shell model \cite{michel23}, and the importance of the $3N$+$N$ structure has been discussed.
}

In Fig. \ref{fig:radius_4}, we plot the radii of protons and neutrons of the $0^+$ states of $^4$He in the GCM calculation
as functions of the excitation energy.
We investigate the behavior of the $0^+_2$ state because this is a resonance.
It is found the ground state shows the smallest values for two kinds of radii and the second state also shows the smallest radii among the excited states.
In particular, the proton radius is more compact than that of other excited states.
This property suggests that the $0^+_2$ state is a resonance and the other excited states are the continuum states with one proton emission from $^4$He.
We also confirm that the continuum states tend to show a small mixing of the individual AMD configurations, which is a different aspect from the resonance case. 

\red{
  The charge radius of the $0^+_2$ state is 4.48 fm as shown in Table \ref{tab:ene_radius1},
  which is much larger than 1.75 fm of the $0^+_1$ state and is comparable to 5.3 fm reported in the four-body calculation \cite{hiyama04}.
}

From these results, the multicool method with the pseudo potential nicely produces the cluster configurations in the excited states of $^4$He without any assumption.
The rigorous treatment of the scattering states is a future problem and the complex scaling can be a promising approach combined with the present method \cite{myo23}.

\subsection{$^5$Li}\label{sec:result5}

We start to discuss the results of the Li isotopes from $^5$Li.
We calculate the unbound nucleus $^5$Li under the bound-state approximation.
The lowest threshold of the particle emission is $\alpha$+$p$ and the second threshold is $^3$He+$d$.
Experimentally, it is known that the $3/2^+$ resonance exists above the $^3$He+$d$ threshold energy by 0.21 MeV with a small decay width of 0.27 MeV \cite{tunl,nndc}.
It is interesting to examine the clustering state in relation to the threshold of $^3$He+$d$ as well as that of $\alpha$+$p$ in the multicool method.
The mirror $3/2^+$ resonance is also observed in $^5$He above the $^3$H+$d$ threshold energy by 0.05 MeV with a small decay width of 0.075 MeV.

\begin{figure}[b]
\centering
\includegraphics[width=5.6cm,clip]{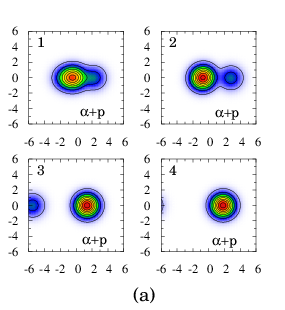}~~~~~~
\includegraphics[width=5.6cm,clip]{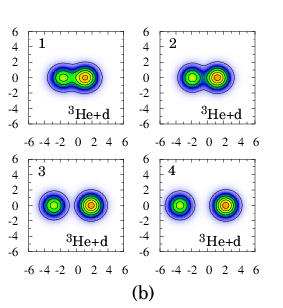}
\caption{
  Intrinsic density distributions of the representative configurations of $^{5}$Li in the spin-fix case for positive parity.
  (a) four panels represent the lowest state. (b) four panels represent the excited state.
  Units of densities and axes are fm$^{-3}$ and fm, respectively.
}
\label{fig:density_5}
\end{figure}

In Fig. \ref{fig:density_5} (a), we show the lowest-state configurations for the positive parity state of $^5$Li in the four panels;
one can confirm the $\alpha$+$p$ structure with various relative distances.
The multicool method nicely describes the superposition of the basis states with the optimal radial distributions.
When we calculate the negative parity state of $^5$Li, a similar $\alpha$+$p$ configuration is obtained.
In the four panels in Fig. \ref{fig:density_5} (b), the configurations of the excited state are shown
for the positive parity state with a pseudo potential.
One confirms the $^3$He+$d$ cluster structure with various relative distances, which could form a resonance.
We superpose these configurations in the GCM calculation.

\begin{figure}[t]
\centering
\includegraphics[width=7.5cm,clip]{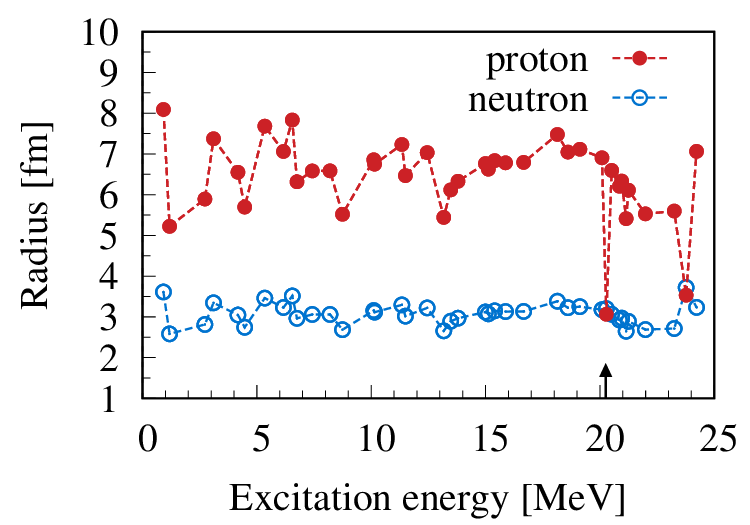}
\caption{
  Radii of protons (red solid circles) and neutrons (blue open circles) of $^5$Li ($3/2^+$) as functions of the excitation energy. Units are fm.
  The arrow indicates the excitation energy corresponding to the candidate of the $3/2^+$ resonance.
}
\label{fig:radius_5}
\end{figure}

\begin{figure}[t]
\centering
\includegraphics[width=7.0cm,clip]{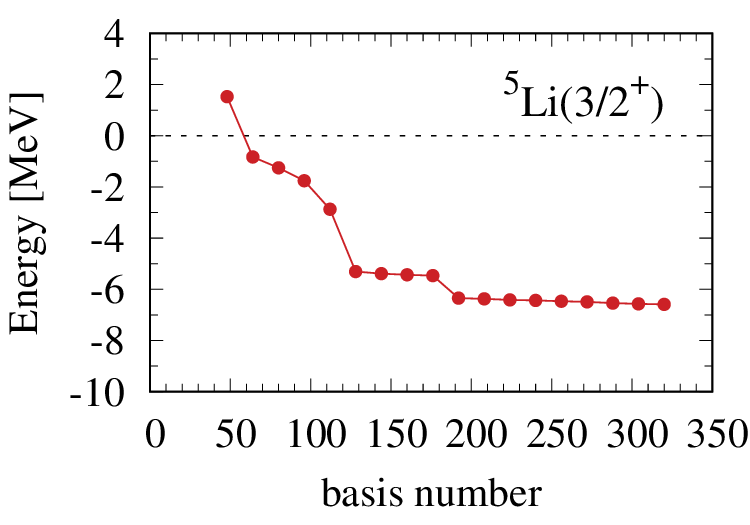}~~~~~
\includegraphics[width=7.0cm,clip]{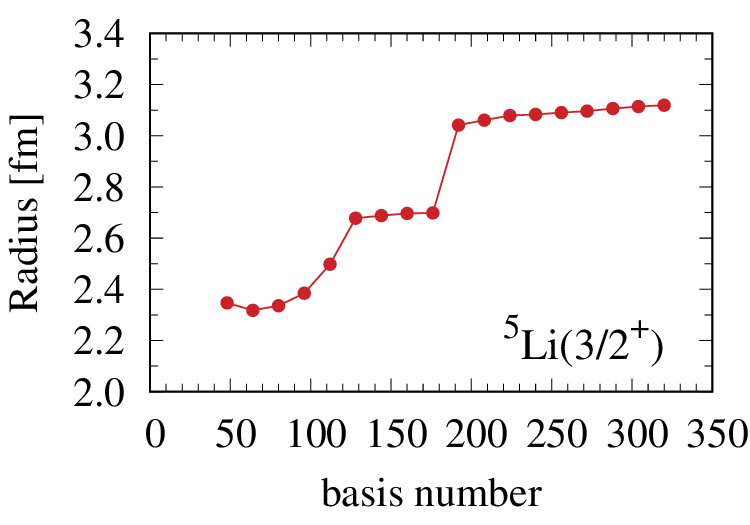}
\caption{\red{
  Convergences of the total energy and matter radius of $^5$Li ($3/2^+$) as functions of the basis number in the GCM calculation. 
  Units are MeV for energy and fm for radius.}
}
\label{fig:cnv_5}
\end{figure}

Experimentally, the $3/2^+$ excited state is observed just above the $^3$He+$d$ threshold energy and the present calculation is consistent with this property.
In Fig. \ref{fig:radius_5}, we show the radius of the $3/2^+$ state of $^5$Li as functions of the excitation energy, similar to the $^4$He case.
It is found that many states are continuum states with a large proton radius, which indicates the continuum states of the $\alpha$+$p$ component.
Among them, around $E_x=20$ MeV, there is a state indicated by the arrow, showing the common values of proton and neutron radii of 3 fm,
which is consistent with the $^3$He+$d$ clustering picture because both clusters involve protons and neutrons. 
In this state, the radii of matter, proton, and neutron are 3.12 fm, 3.06 fm, and 3.20 fm, respectively.
The dominant configuration of this state certainly shows a $^3$He+$d$ cluster structure in the AMD configuration and the highest weight is 79\%.
The total energy of this state is $-6.59$ MeV, which is close to the threshold energy of $-6.76$ MeV for the separation of $^5$Li into $^3$He and $d$ assuming their single AMD configurations.
These situations are consistent with the experimental property of the $3/2^+$ state of $^5$Li
and hence the present $3/2^+$ state can be regarded as the clustering state consisting of $^3$He and $d$.

\red{
It is noted that the present $3/2^+$ state is obtained with the bound-state approximation.
We examine the stability of the solution of the $3/2^+$ state as functions of the GCM basis number in Fig. \ref{fig:cnv_5},
and one can confirm the convergences of the energy and radius of the state.}

In $^5$He, the mirror system of $^5$Li, we also confirm the corresponding mirror state of $3/2^+$ at $-7.67$ MeV of the total energy,
which is close to the $^3$H+$d$ threshold energy of $-7.55$ MeV with the single AMD configurations for both clusters.
Experimentally, the $3/2^+$ resonance of $^5$He is located just above the $^3$H+$d$ threshold energy
and the present calculation is consistent with the experimental situation as well as the results of $^5$Li.
For this state, the $^3$H+$d$ is a dominant configuration and the radii of matter, proton, and neutron are 3.15 fm, 3.20 fm, and 3.12 fm, respectively.
The dominant AMD configuration of this state shows a $^3$H+$d$ structure with the highest weight of 79\%.
The isospin symmetry is well confirmed in the $3/2^+$ resonance of $A=5$.
\red{
  The discussion on the $^3$H+$d$ structure in the excited state of $^5$He is also
  shown in the tensor-optimized AMD using the realistic AV8$^\prime$ potential \cite{myo21}.
}
From these analyses on the excited states of the $A=5$ mirror systems in the multicool method,
we predict the existence of the excited cluster states consisting of a $3N$+$d$ structure near the corresponding threshold energies.

\subsection{$^6$Li}\label{sec:result6}

\begin{figure}[b]
\centering
\includegraphics[width=15.3cm,clip]{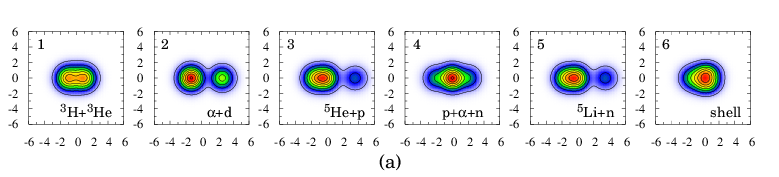}\vspace*{-0.2cm}\\  
\includegraphics[width=15.3cm,clip]{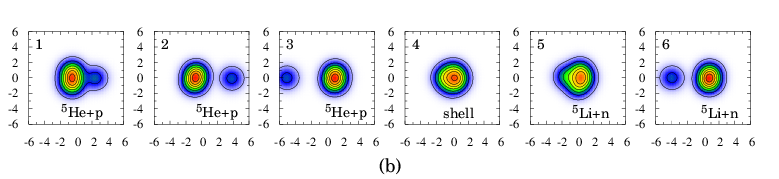}    
\caption{
  Intrinsic density distributions of the representative configurations of $^{6}$Li in the spin-fix case for positive parity.
  (a) upper six panels: the ground state.
  (b) lower six panels: the excited state.
  Units of densities and axes are fm$^{-3}$ and fm, respectively.
}
\label{fig:density_6}
\end{figure}

We discuss $^6$Li.
In Fig. \ref{fig:density_6}, we show the intrinsic density distributions of the representative configurations of $^6$Li for positive parity state in the multicool method.
The six panels in (a) are for the ground state and the six panels in (b) are for the excited state with the pseudo potential.
In the ground state, the spatially compact $^3$H+$^3$He, and $\alpha$+$d$, $^5$He+$p$, $^5$Li+$n$, $\alpha$+$p$+$n$, and shell-like configurations are obtained.
In the excited state, $^5$He+$p$, $^5$Li+$n$, and the compact shell-like configurations are confirmed and the $5N$+$N$ cluster configurations show the various relative distances.

In the experiments, the threshold energies of these cluster configurations are located near the low-excitation energy region,
such as the $\alpha$+$p$+$n$ being 3.7 MeV.
In the present calculation, these cluster configurations become an important ingredient to describe the $^6$Li structure as well as the shell-like one.
Here, the $^3$H+$^3$He threshold energy is highly located at 15.8 MeV measured from the ground state, but
we confirm this configuration in the ground state with a small relative distance.
We also confirm the compact $^3$H+$^3$H configuration in $^6$He in the multicool method \cite{myo23b},
and the importance of this configuration has been discussed in $^6$He with the microscopic cluster models \cite{csoto93,arai99,aoyama06}.
In the multicool method, the compact $3N$+$3N$ configuration is automatically involved to describe the structures of the $A=6$ system. 

\begin{figure}[t]
\centering
\includegraphics[width=4.8cm,clip]{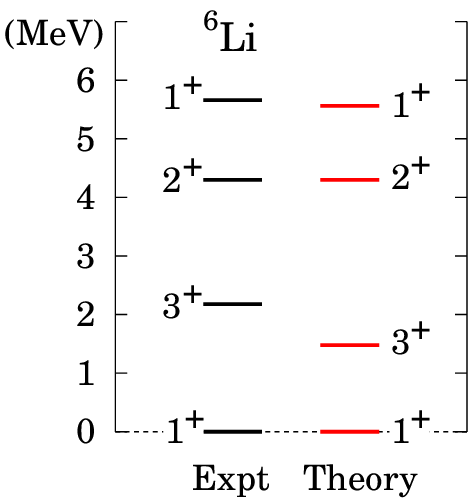}
\caption{Energy levels of $^6$Li in the multicool method in comparison with the experiments.}
\label{fig:ene_6}
\end{figure}

We show the energy spectrum with the GCM calculation in Fig. \ref{fig:ene_6}, which reproduces the experimental levels of the $T=0$ state.
In the ground $1^+$ state, the mixing of the configurations of $^3$H+$^3$He and $\alpha$+$d$ shown in Fig. \ref{fig:density_6} (a) is 55\% and 84\%, respectively. 
We do not show the excited energy levels with $T=1$, because we do not impose the constraint on the isospin of the total system in the present multicool method. 
The variation in the $T=1$ space is interesting to investigate the isospin effect, which is a future work.

\red{
There is a six-body calculation of $^6$Li in the few-body method with the correlated Gaussian functions \cite{satsuka19}.
They evaluate the $E1$ transition strength of $^6$Li corresponding to the photoabsorption experiment
and discuss the effect of cluster configurations of $\alpha$+$p$+$n$ and $^3$H+$^3$He.
Similarly, it would be interesting to investigate the roles of various cluster configurations in the nuclear excitations using the present multicool method.
}

\subsection{$^7$Li}\label{sec:result7}

\begin{figure}[t]
\centering
\includegraphics[width=13.2cm,clip]{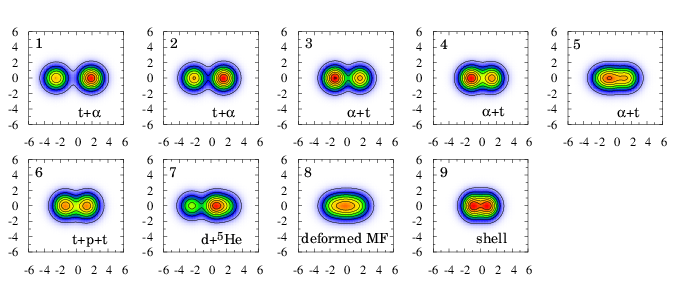} 
\caption{
  Intrinsic density distributions of the representative configurations of the ground state of $^7$Li in the spin-fix case.
  Units of densities and axes are fm$^{-3}$ and fm, respectively.
}
\label{fig:density_7GS}
\end{figure}

\begin{figure}[th]
\centering
\includegraphics[width=13.2cm,clip]{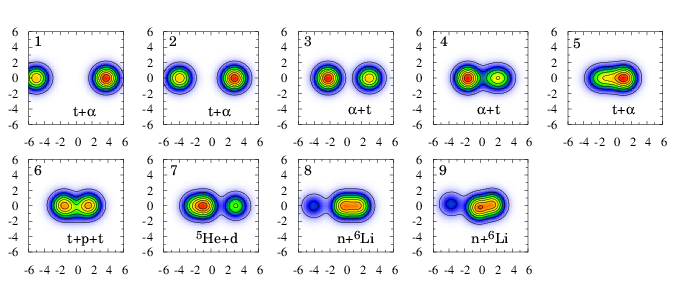} 
\caption{
  The same as Fig. \ref{fig:density_7GS}, but for the excited state of $^7$Li with the pseudo potential.
}
\label{fig:density_7EX}
\end{figure}

We discuss $^7$Li.
In Figs. \ref{fig:density_7GS} and \ref{fig:density_7EX}, we show the intrinsic density distributions of the representative configurations of $^7$Li
for negative parity in the multicool method.
There are nine panels for the ground state and the excited state, respectively.
In the ground state in Fig. \ref{fig:density_7GS}, the $\alpha$+$t$ configuration is mainly obtained with various relative distances between clusters.
In addition to them, $t$+$p$+$t$, $^5$He+$d$, deformed mean-field (MF), and compact shell-like configurations are obtained.
In the excited state in Fig. \ref{fig:density_7EX}, $\alpha$+$t$, $^5$He+$d$, and $t$+$p$+$t$ configurations can be confirmed with a large relative distance between clusters,
and the $^6$Li+$n$ configuration is newly obtained with the elongated distribution to the horizontal axis.

These cluster configurations are related to the corresponding thresholds appearing in the low-excitation energy region of $^7$Li.
Experimentally, the corresponding threshold energies are located up to around 10 MeV of the excitation energy
such as 2.5 MeV ($\alpha$+$t$), 7.2 MeV ($^6$Li+$n$), and 9.5 MeV ($^5$He+$d$).

\begin{figure}[th]
\centering
\includegraphics[width=5.2cm,clip]{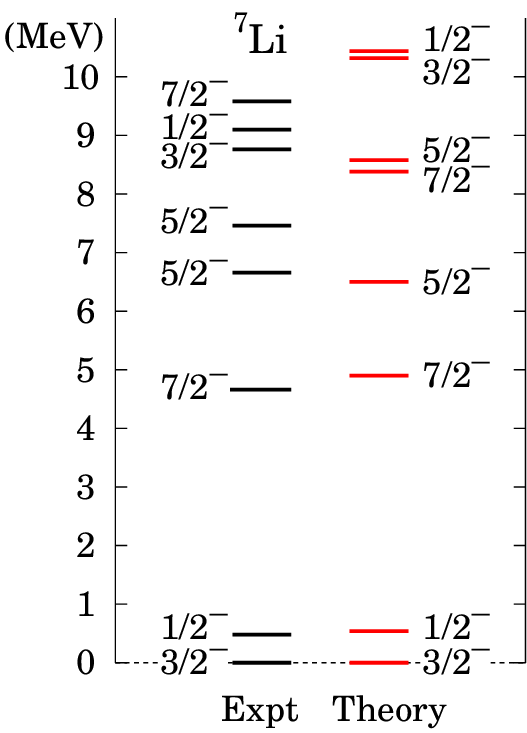}
\caption{Energy levels of $^7$Li in the multicool method in comparison with the experiments.}
\label{fig:ene_7}
\end{figure}

We show the energy spectrum with the GCM calculation in Fig. \ref{fig:ene_7}, which reproduces the experimental levels.
In the ground $3/2^-$ state, among the configurations shown in Fig. \ref{fig:density_7GS}, 
the mixing of the 3rd $\alpha$+$t$ configuration is the largest value of 85\%.
The $t$+$p$+$t$ configuration (6th) is 68\% and the $d$+$^5$He configuration (7th) is 44\%.


In the excited states of $^7$Li, it is found that four states ($1/2^-_2$, $3/2^-_2$, $5/2^-_2$, and $7/2^-_2$)
above the excitation energies of 8 MeV commonly show very large radii of around 4.5 fm shown in Table \ref{tab:monopole_7} with the values of $R_{\rm F}$.
In these states, the dominant configuration is an $\alpha$+$t$ structure with a large relative distance, such as the 1st panel in Fig.~\ref{fig:density_7EX}.
Hence these states can be regarded as the well-developed $\alpha$+$t$ cluster states.

It is known that cluster states tend to give a large monopole transition strength, which can be a signature to prove the clustering in the states \cite{yamada08}.
According to this discussion, we calculate the monopole transition strengths of the corresponding four states of $^7$Li from the lower-energy states,
which are summarized in Table \ref{tab:monopole_7}.
In the table, IS0 and $E0$ are the monopole transitions of the isoscalar and electric (proton) parts, respectively,
using the operators of the squared radius.
We also evaluate the transition of only the neutron part for reference \cite{ito14}.
Most of the strengths show large values in comparison with the single-particle strength of $\sqrt{5/(8\nu^2)}=3.4$ fm$^2$ estimated from the $0p$ to $1p$ orbits \cite{ito14}.
These results support the spatial extension due to the $\alpha$+$t$ clustering of $^7$Li.

\begin{table}[t]
\begin{center}
\caption{
  Monopole transition strengths of $^7$Li with the isoscalar (IS0), electric ($E0$, proton), and neutron parts.
  Units are fm$^2$. The matter radii of the initial ($R_{\rm I}$) and final ($R_{\rm F}$) states are also shown in units of fm.
}
\label{tab:monopole_7}
\renewcommand{\arraystretch}{1.1}
\begin{tabular}{llllllllllll}
\noalign{\hrule height 0.5pt}
                      && IS0~~&& $E0$~~&& Neutron~~&& $R_{\rm I}$~~&& $R_{\rm F}$~~~\\
\noalign{\hrule height 0.5pt}
$1/2^-_1 \to 1/2^-_2$ && ~~9.05 && ~3.73  && ~5.33  && 2.45  &&  4.34 \\
$3/2^-_1 \to 3/2^-_2$ && ~~8.99 && ~3.75  && ~5.24  && 2.39  &&  4.30 \\
$5/2^-_1 \to 5/2^-_2$ && 34.91  && 14.28  && 20.63  && 2.82  &&  4.57 \\
$7/2^-_1 \to 7/2^-_2$ && 16.31  && ~6.73  && ~9.57  && 2.36  &&  4.61 \\
\noalign{\hrule height 0.5pt}
\end{tabular}
\end{center}
\end{table}

\subsection{$^8$Li}\label{sec:result8}

\begin{figure}[t]
\centering
\includegraphics[width=15.3cm,clip]{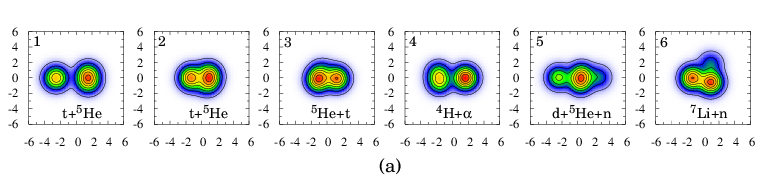}\vspace*{-0.2cm}\\   
\includegraphics[width=15.3cm,clip]{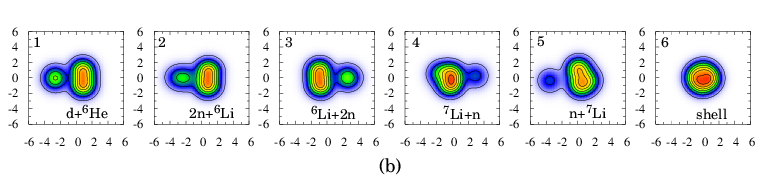}     
\caption{
  Intrinsic density distributions of the representative configurations of $^8$Li for (a) upper six panels: the ground state, and
  (b) lower six panels: the excited state, in the spin-fix case with positive parity.
  Units of densities and axes are fm$^{-3}$ and fm, respectively.
}
\label{fig:density_8}
\end{figure}

We discuss $^8$Li.
In Fig. \ref{fig:density_8}, we show the intrinsic density distributions of the representative configurations of $^8$Li for the positive parity in the multicool method; the six panels in (a) are for the ground state, and the six panels in (b) are for the excited state.
In the ground-state configurations, $^5$He+$t$, $\alpha$+$^4$H, $^5$He+$d$+$n$, and $^7$Li+$n$ are obtained.
The $^5$He+$t$ configurations show various relative distances between clusters.

In the excited-state configurations, $^6$He+$d$, $^6$Li+$2n$, $^7$Li+$n$, and the compact shell-like state are obtained,
in which the last neutron in the $^7$Li+$n$ configuration is well separated from $^7$Li in comparison with the ground state case.

These cluster configurations are related to the thresholds of the corresponding cluster emissions,
which are observed experimentally up to about 10 MeV of the excitation energy
such as $^7$Li+$n$ (2.0 MeV), $^5$He+$t$ (5.3 MeV), $^6$Li+$2n$ (9.3 MeV), and $^6$He+$d$ (9.8 MeV).
This property is commonly confirmed in the results of $^6$Li and $^7$Li in the multicool method.

\begin{figure}[t]
\centering
\begin{minipage}[t]{0.36\textwidth}
    \includegraphics[width=4.65cm,clip]{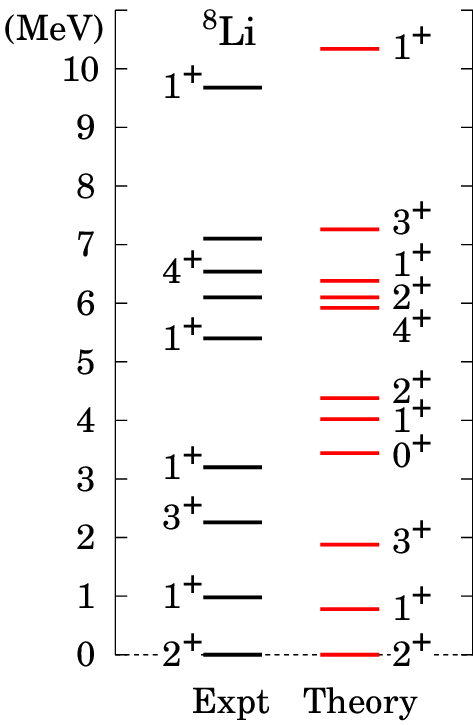}
    \caption{Energy levels of $^8$Li in the multicool method in comparison with the experiments.}
    \label{fig:ene_8}
\end{minipage}~~~~
\begin{minipage}[t]{0.57\textwidth}
    \includegraphics[width=8.70cm,clip]{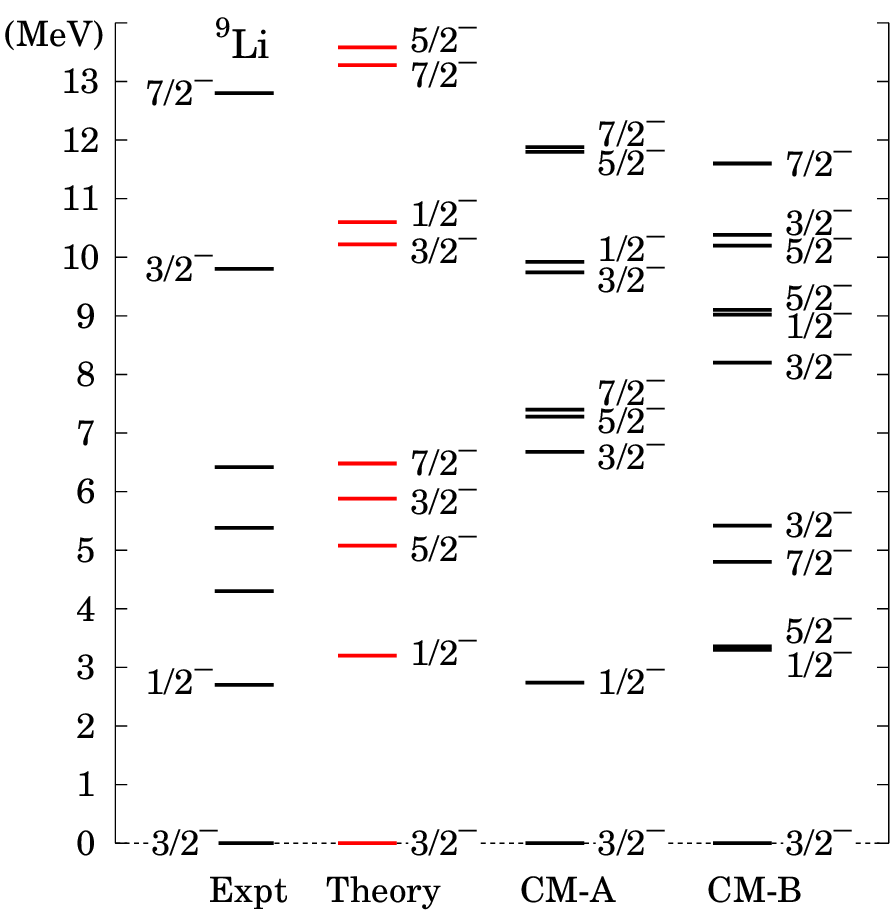}  
    \caption{Energy levels of $^9$Li in the multicool method (Theory) in comparison with the experiments.
      The experiments of the higher two levels are taken from Ref. \cite{ma21}.
      The results of other cluster models; CM-A \cite{kanada12} and CM-B \cite{furumoto13} are also shown.
    }
    \label{fig:ene_9}
\end{minipage}
\end{figure}

We show the energy spectrum with the GCM calculation in Fig. \ref{fig:ene_8}, which reproduces the experimental levels.
We also predict some levels such as $0^+$, $2^+_2$, $2^+_3$, and $3^+_2$, which are not confirmed experimentally,
but also predicted in the ab initio Green's function Monte Carlo \cite{carlson15} and the shell model calculation including the tensor correlation from the nucleon-nucleon interaction \cite{myo12}.

\subsection{$^9$Li}\label{sec:result9}

Finally, we discuss $^9$Li.
In Figs. \ref{fig:density_9_demo}, \ref{fig:density_9_pse500}, \ref{fig:density_9_pse800}, and \ref{fig:density_9_pse1600},
we already discuss the intrinsic density distributions of the representative configurations of $^9$Li in the ground and excited states.
We confirm the various cluster configurations including the linear-chain structure as well as the compact shell-like configuration.
The cluster configurations are related to the presence of the thresholds of the corresponding cluster emissions,
which is a common feature seen in other Li isotopes in the multicool method.

We show the energy spectrum with the GCM calculation in Fig. \ref{fig:ene_9}, which reproduces the observed experimental levels and is consistent with the results of other cluster models of $^9$Li \cite{furumoto13,kanada12,kanada16}.
We predict four levels of $1/2^-_2$, $3/2^-_3$, $5/2^-_2$, and $7/2^-_2$ above the excitation energy of 10 MeV.
We confirmed that these four states commonly have the large amplitudes of the linear-chain configurations shown in Fig. \ref{fig:density_9_pse800},
with maximally about 60--70\% of the single AMD configuration for $1/2^-_2$, $5/2^-_2$, and $7/2^-_2$, and about 40\% for $3/2^-_3$,
which are larger than those of other configurations.
These states also have a large radius of around 3 fm as shown in Table \ref{tab:9Li_linear_chain}, and
the neutron radii are larger than the proton radii for four states.
The $3/2^-_3$ state relatively shows the small radii among them due to the smaller mixing of linear-chain configurations
than those of the other three states.
From these properties, we can assign four states the candidate of a linear-chain state,
which is fragile because of the mixing of the various kinds of cluster configurations shown in Fig. \ref{fig:density_9_pse800}.
This property is similar to the $0^+_2$ state of $^{10}$Be \cite{myo23b}.

In the recent experiment \cite{ma21}, there is a report of the candidates of $3/2^-$ and $7/2^-$ states at the excitation energies of 9.8 MeV and 12.8 MeV,
respectively, which agree with our predictions of 10.22 MeV and 13.28 MeV, respectively.
In the experiment, the isoscalar monopole transition from $3/2^-_1$ to $3/2^-_3$ is reported as 8.1(8) fm$^2$,
while the calculated value is obtained as 3.57 fm$^2$,
which is smaller than the experimental one, but gives the same order of the magnitude.
The present value is also comparable to the single particle strength of 3.4 fm$^2$ from the $0p$ orbit to the $1p$ orbit \cite{ito14}
as discussed in $^7$Li.

On the other hand, the transition from $3/2^-_1$ to $3/2^-_2$ is 0.42 fm$^2$ in the calculation,
which is small and a common feature of the value reported in Ref.~\cite{furumoto13}.
In the shell model calculation including the tensor correlation \cite{myo12},
the main configuration of $3/2^-_2$ is the pairing excitation of neutrons from $p_{3/2}$ to $p_{1/2}$ orbits,
which can reduce the monopole transition from the ground state due to the spin condition of the single particle.
The matter radius of $3/2^-_2$ is 2.42 fm, which is smaller than 2.63 fm of $3/2^-_3$.
These transitions are summarized in Table \ref{tab:monopole_9} together with those of the electric and neutron parts.
For reference, we show the transitions from $1/2^-_1$ to $1/2^-_2$ having linear-chain configurations,
and the tendency of the transitions is found to be similar to that from $3/2^-_1$ to $3/2^-_3$.

In the summary of the results of Li isotopes in the multicool method,
we confirm various cluster configurations associated with the corresponding thresholds in the low-excitation energy region.
These results would be useful to examine the threshold rule for unstable nuclei.
The detailed analysis of the structures of each of Li isotopes is in progress. 

\begin{table}[t]
\begin{center}
  \caption{
    Radii of matter, proton, and neutron ($r_{\rm m}$, $r_{\rm p}$, and $r_{\rm n}$) of $1/2^-_2$, $3/2^-_3$, $5/2^-_2$, and $7/2^-_2$ in $^9$Li,
    which have the large weight of the linear-chain configurations shown in Fig. \ref{fig:density_9_pse800}.
    Units are fm.}
\label{tab:9Li_linear_chain}
\renewcommand{\arraystretch}{1.1}
\begin{tabular}{llllllllllll}
\noalign{\hrule height 0.5pt}
          &&~~$r_{\rm m}$~~&&~~$r_{\rm p}$~~&&~~$r_{\rm n}$~~\\
\noalign{\hrule height 0.5pt}
$1/2^-_2$ &&  2.92 &&  2.59  &&  3.07   \\
$3/2^-_3$ &&  2.63 &&  2.34  &&  2.77   \\
$5/2^-_2$ &&  2.97 &&  2.64  &&  3.12   \\
$7/2^-_2$ &&  2.85 &&  2.52  &&  3.00   \\
\noalign{\hrule height 0.5pt}
\end{tabular}
\end{center}
\end{table}
\begin{table}[t]
\begin{center}
  \caption{
    Monopole transition strengths of $^9$Li from $1/2^-_1$ to $1/2^-_2$ and from $3/2^-_1$ to $3/2^-_{2,3}$
    for the isoscalar (IS0), electric ($E0$, proton), and neutron parts in units of fm$^2$.
    The values with square brackets are the experimental data \cite{ma21}.
    The matter radii of the initial ($R_{\rm I}$) and final ($R_{\rm F}$) states are also shown in units of fm.
  }
\label{tab:monopole_9}
\renewcommand{\arraystretch}{1.1}
\begin{tabular}{llllllllllll}
\noalign{\hrule height 0.5pt}
                      && IS0~~          && $E0$~~&& Neutron~~&& $R_{\rm I}$~~&& $R_{\rm F}$~~~\\
\noalign{\hrule height 0.5pt}
$1/2^-_1 \to 1/2^-_2$ &&  4.27          && 1.11  && 3.17  &&  2.39  && 2.92 \\
$3/2^-_1 \to 3/2^-_2$ &&  0.42          && 0.07  && 0.35  &&  2.29  && 2.42 \\
$3/2^-_1 \to 3/2^-_3$ &&  3.57 [8.1(8)] && 0.98  && 2.58  &&  2.29  && 2.63 \\
\noalign{\hrule height 0.5pt}
\end{tabular}
\end{center}
\end{table}

\section{Summary}\label{sec:summary}

We developed the variational method of the multi-Slater determinants based on the antisymmetrized molecular dynamics (AMD) for nuclei.
In this method, we obtain the configurations of the ground state, each of which is optimized simultaneously under the minimization of the total energy.
For the excited states, we impose the orthogonal condition to the ground-state configurations.
We call this framework the ``multicool method'' because we extend the cooling equation of the single AMD basis state to the multi-AMD basis case.
The multicool method has the advantage of describing various kinds of possible cluster configurations in a nucleus without a priori knowledge,
which is done in $^{10}$Be \cite{myo23b}.

In this study, we focus on the Li isotopes, in which the thresholds of the various kinds of cluster emissions exist in the low-excitation energy region.
Hence the corresponding cluster configurations are expected to contribute to the structures of Li isotopes, similar to $^{10}$Be.

The ground-state and excited-state configurations of Li isotopes are determined in the multicool method
and we discuss the characteristic properties of the configurations from the density distributions.
Variety of the cluster configurations are confirmed as well as the shell-like one and
this property is related to the existence of the thresholds of the associated cluster emissions.
Finally, we superpose all the AMD basis states in the GCM calculation and discuss the energy levels.

For $^4$He, which often appears in Li isotopes as an $\alpha$ cluster,
the $0s$-shell configuration is dominated in the ground $0^+$ state and the $3N$+$N$ configuration gives the partial contributions.
In the $0^+_2$ state, $^3$H+$p$ configuration becomes dominant and the state is obtained near the corresponding threshold energy.  

For $^5$Li, we focus on the excited $3/2^+$ state,
which is dominated by the $^3$He+$d$ cluster configuration and is obtained near the threshold energy of the corresponding clusters.
In the mirror state of $^5$He, the same tendency is confirmed with the $^3$H+$d$ clusters.
In the $A=5$ system, the emergence of the $3N$+$2N$ cluster configuration near the corresponding threshold energy
is consistent with the experimental situation of the $3/2^+$ states.

For $^6$Li, the $\alpha$+$d$ configuration becomes dominant in the ground state, but the compact $^3$H+$^3$He configuration can also be mixed in the state.
For $^7$Li, $6N$+$N$, $5N$+$2N$, $4N$+$3N$, $3N$+$3N$+$N$ and the compact shell-like configurations are obtained.
We confirm that the excited states above 8 MeV of the excitation energy are dominated by the $\alpha$+$t$ configuration
with a large relative distance between clusters.
These states commonly show large radii and large monopole transitions beyond the single-particle estimation,
which support the $\alpha$+$t$ clustering in $^7$Li.

For $^8$Li, $7N$+$N$, $6N$+$2N$, $5N$+$3N$ and the compact shell-like configurations are confirmed.
We predict some of the energy levels, which are consistent with the shell-model calculation including the tensor correlation \cite{myo12}.

For $^9$Li, $8N$+$N$, $7N$+$2N$, $6N$+$3N$, $5N$+$4N$ and the shell-like configurations are confirmed. 
In the excited states, we predict the linear-chain configurations, in which various sub-clusters are involved.
This property is similar to the $0^+_2$ state of $^{10}$Be, the isotone of $^9$Li.
The four states above 10 MeV of the excitation energy have large components of the linear-chain configurations with large radii.
The excitation energies of the $3/2^-_3$ and $7/2^-_2$ states agree with the experiments.
We predict the relatively large monopole strength to the $3/2^-_3$ state, the order of which agrees with the experimental value.

From the results of Li isotopes, we confirm the validity of the multicool method to obtain the optimal AMD configurations for nuclear structure.
Various cluster configurations are obtained automatically and even the unbound nuclei such as $^4$H and $^5$He can be a constituent of a clustering system.
The appearance of the cluster configurations can be related to the corresponding thresholds for cluster emissions.

In the forthcoming study, we perform a detailed analysis of the structures of Li isotopes.
We also consider applying the multicool method to heavier mass nuclei such as $^{12}$C \cite{cheng24}, $^{16}$O, and the $sd$-shell nuclei.
The application to the ab initio type calculations based on AMD is in progress \cite{myo17a,lyu20}.

\section*{Acknowledgments}
This work was supported by JSPS KAKENHI Grant Numbers JP21K03544 and JP22K03643, JST ERATO Grant No. JPMJER2304, Japan, and JSPS A3 Foresight Program.
M.L. acknowledges support from the National Natural Science Foundation of China (Grants No. 12105141),
and the Jiangsu Provincial Natural Science Foundation (Grants No. BK20210277).

\nc\PTEP[1]{Prog.\ Theor.\ Exp.\ Phys.,\ \andvol{#1}} 
\nc\PPNP[1]{Prog.\ Part.\ Nucl.\ Phys.,\ \andvol{#1}} 



\begin{thebibliography}{00}

\bibitem{ikeda68}    K. Ikeda, N. Takigawa, and H. Horiuchi, \PTPS{E68,464,1968} 
\bibitem{horiuchi12} H. Horiuchi, K. Ikeda, and K. Kat\=o, \PTPS{192,1,2012} 
\bibitem{freer18}    M. Freer, H. Horiuchi, Y. Kanada-En'yo, D. Lee, and Ulf-G. Mei{\ss}ner,  \JL{Rev. Mod. Phys.,90,035004,2018}. 
  
\bibitem{kanada03}    Y. Kanada-En'yo, M. Kimura, and H. Horiuchi, \JL{Compt. Rendus Phys.,4,497,2003}. 
\bibitem{kimura16}    M.~Kimura, T.~Suhara and Y.~Kanada-En'yo, \JL{Eur. Phys. J. A,52,373,2016}. 

\bibitem{myo23b} T.~Myo, M. Lyu, Q. Zhao, M. Isaka, N. Wan, H. Takemoto, and H.~Horiuchi, \PRC{108,064314,2023} 

\bibitem{tilley04}  D.~R.~Tilley, J.~H.~Kelley, J.~L.~Godwin, D.~J.~Millener, J.~E.~Purcell, C.~G.~Sheu and H.~R.~Weller,
  \NPA{745,155,2004} 

\bibitem{tunl}        https://nucldata.tunl.duke.edu/nucldata/index.shtml/ . 
\bibitem{nndc}        https://www.nndc.bnl.gov/nudat3/ . 
\bibitem{faessler71} A. Faessler, K. W. Schmid, and  A. Plastino,  \NPA{174,26,1971}  
  
\bibitem{ogawa11} Y. Ogawa and H. Toki, \JL{Annals of Physics,326,209,2011}. 

\bibitem{pillet17} N.~Pillet, C.~Robin, M.~Dupuis, G.~Hupin and J.~F.~Berger, \JL{Eur. Phys. J. A,53,49,2017} and the references therein. 

\bibitem{matsumoto23} M. Matsumoto, Y. Tanimura, K. Hagino, \PRC{108,L051302,2023}  
\red{
\bibitem{shimizu22}  N. Shimizu, \JL{Physics,2022,1081,2022}, and the references therein.  
}
  
\bibitem{tian24} J.~Tian, Z.~Cheng, C.~Yu, M.~Lyu, T.~Myo, M.~Isaka, H.~Toki, H.~Horiuchi, A.~Dot\'e and H.~Takemoto, \textit{et al.}, \PLB{855,138816,2024}  

\bibitem{cheng24} Z.~Cheng, M.~Lyu, T.~Myo, H.~Horiuchi, H.~Toki, Z.~Ren, M.~Isaka, M.~Mao, H.~Takemoto and N.~Wan, \textit{et al.}, e-Print: 2406.15060 [nucl-th] 
  
  
\bibitem{itagaki00}   N. Itagaki and S. Okabe, \PRC{61,044306,2000}  
\bibitem{ito06}       M.~Ito, \PLB{636,293,2006} 
\bibitem{suhara10}    T.~Suhara and Y.~Kanada-En'yo, \PTP{123,303,2010} 

 
\bibitem{aoyama01}   S. Aoyama, K. Kat\=o, and K. Ikeda, \PTPS{142,35,2001} 
\bibitem{myo14}      T. Myo, Y. Kikuchi, H. Masui, and K. Kat\=o, \PPNP{79,1,2014}  


\bibitem{varga95}   K. Varga, Y. Suzuki, and I. Tanihata \PRC{52,3013,1995} 


\bibitem{furumoto13} T.~Furumoto, T.~Suhara and N.~Itagaki, \PRC{87,064320,2013};\PRC{90,039902(E),2014} 
  

\bibitem{sick08} I. Sick, \PRC{77,041302(R),2008}  
  
\bibitem{mueller07} P.~Mueller, I.~A.~Sulai, A.~C.~C.~Villari, J.~A.~Alc\'antara-N\'u\~nez, R.~Alves-Cond\'e, K.~Bailey,
  G.~W.~F.~Drake, M.~Dubois, C.~El\'eon, G.~Gaubert, \textit{et al.}, \PRL{99,252501,2007}  

\bibitem{nortershauser11} W.~N\"ortersh\"auser, T. Neff, R. S\'anchez, and I. Sick, \PRC{84,024307,2011}  
\red{
\bibitem{hiyama04}   E. Hiyama, B. F. Gibson, and M. Kamimura, \PRC{70,031001(R),2004}  
\bibitem{horiuchi08} W. Horiuchi and Y. Suzuki, \PRC{78,034305,2008}  
\bibitem{michel23}   N. Michel, W. Nazarewicz, M. P{\l}oszajczak, \PRL{131,242502,2023};\PRL{133,239901,2024}     
}
\bibitem{myo23} T.~Myo and H.~Takemoto, \PRC{107,064308,2023} 
\red{
\bibitem{myo21}      T. Myo, M. Lyu, H. Toki, and H. Horiuchi, \PTEP{2021,023D043,2021}   
}
\bibitem{csoto93} A. C\'sot\'o, \PRC{48,165,1993}

\bibitem{arai99} K.~Arai, Y.~Suzuki and R.~G.~Lovas, \PRC{59,1432,1999}

\bibitem{aoyama06} S.~Aoyama, N.~Itagaki and M.~Oi, \PRC{74,017307,2006}
\red{
\bibitem{satsuka19}  S. Satsuka and W. Horiuchi, \PRC{100,02433,2019}  
}
\bibitem{yamada08} T.~Yamada, Y.~Funaki  H.~Horiuchi, K.~Ikeda, and A.~Tohsaki, \PTP{120,1139,2008}  

\bibitem{ito14} M.~Ito and K.~Ikeda, \JL{Rep. Prog. Phys.,77,096301,2014}.  

\bibitem{carlson15} J.~Carlson, S.~Gandolfi, F.~Pederiva, S.~C.~Pieper, R.~Schiavilla, K.~E.~Schmidt and R.~B.~Wiringa,
  \JL{Rev. Mod. Phys.,87,1067,2015}. 

\bibitem{myo12}  T. Myo, A. Umeya, H. Toki and K. Ikeda, \PRC{86,024318,2012} 

\bibitem{kanada12}   Y. Kanada-En'yo, and T. Suhara, \PRC{85,024303,2012} 
\bibitem{kanada16}   Y. Kanada-En'yo, \PRC{94,024326,2016} 

\bibitem{ma21}  W.~H.~Ma, J.~S.~Wang, D.~Patel, R.~F.~Chen, Y.~Y.~Yang, J.~B.~Ma, S.~L.~Jin, P.~Ma, Q.~Hu and Y.~S.~Song, \textit{et al.}, \PRC{103,L061302,2021} 

\bibitem{myo17a}      T. Myo, H. Toki, K. Ikeda, H. Horiuchi, and T. Suhara, \PLB{769,213,2017}   
\bibitem{lyu20}       M.~Lyu, T.~Myo, H.~Toki, H.~Horiuchi, C.~Xu and N.~Wan, \PLB{805,135421,2020}  


\end{thebibliography}
\end{document}